\documentclass[sn-aps,Numbered,iicol]{sn-jnl}
\usepackage{graphicx}%
\usepackage{multirow}%
\usepackage{amsmath,amssymb,amsfonts}%
\usepackage{amsthm}%
\usepackage{mathrsfs}%
\usepackage[title]{appendix}%
\usepackage{xcolor}%
\usepackage{textcomp}%
\usepackage{manyfoot}%
\usepackage{booktabs}%
\usepackage{algorithm}%
\usepackage{algorithmicx}%
\usepackage{algpseudocode}%
\usepackage{listings}%

\usepackage{ulem}

\newcommand{\nn}{\nonumber}

\begin{document}


\title{Relativistic BGK hydrodynamics}


\author*[1,2]{\fnm{Pracheta}\sur{Singha}}\email{pracheta.singha@gmail.com}
\author*[1,3]{\fnm{Samapan}\sur{Bhadury}}\email{samapan.bhadury@uj.edu.pl}
\author*[1,4,5]{\fnm{Arghya}\sur{Mukherjee}}
\email{arbp.phy@gmail.com}
\author*[1]{\fnm{Amaresh}\sur{Jaiswal}}
\email{a.jaiswal@niser.ac.in}

\affil[1]{School of Physical Sciences, National Institute of Science Education and Research, An OCC of Homi Bhabha National Institute, Jatni 752050, Odisha, India}
\affil[2]{Department of Physics, West University of Timisoara, Bd. Vasile Pârvan 4, Timisoara 300223, Romania}
\affil[3]{Institute of Theoretical Physics, Jagiellonian University, ul. St. \L ojasiewicza 11, 30-348 Krakow, Poland}
\affil[4]{Department of Physics and Astronomy, Brandon University, Brandon, Manitoba R7A 6A9, Canada}
\affil[5]{Ramakrishna Mission Residential College (Autonomous), Narendrapur, Kolkata 700103, India}


\date{\today}


\abstract{
Bhatnagar-Gross-Krook (BGK) collision kernel is employed in the Boltzmann equation to formulate relativistic dissipative hydrodynamics. In this formulation, we find that there remains freedom of choosing a matching condition that affects the scalar transport in the system. We also propose a new collision kernel which, unlike BGK collision kernel, is valid in the limit of zero chemical potential and derive relativistic first-order dissipative hydrodynamics using it. We study the effects of this new formulation on the coefficient of bulk viscosity.
}


	


\maketitle

%
\section{Introduction}

Relativistic Boltzmann equation governs the space-time evolution of the single particle phase-space distribution function of a relativistic system. Moreover, suitable moments of the Boltzmann equation are capable of describing the collective dynamics of the system. Therefore, it has been extensively used to derive equations of relativistic dissipative hydrodynamics and obtain expressions for the transport coefficients \cite{Muller:1967zza, Chapman:1970, Israel:1979wp, Muronga:2006zx, York:2008rr, Betz:2008me, Romatschke:2009im, Denicol:2010xn, Denicol:2012cn, Jaiswal:2014isa, Jaiswal:2015mxa, Gabbana:2017uvc, Blaizot:2017lht, Jaiswal:2022udf, Jaiswal:2020hvk}. The collision term in the Boltzmann equation, which describes change in the phase-space distribution due to the collisions of particles, makes it a complicated integro-differential equation. In order to circumvent this issue, several approximations have been suggested to simplify the collision term in the linearized regime \cite{Bhatnagar:1954zz, Welander, marle1969, anderson1974relativistic, Rocha:2021zcw}.

Bhatnagar-Gross-Krook \cite{Bhatnagar:1954zz}, and independently Welander \cite{Welander}, proposed a relaxation type model for the collision term, which is commonly known as the BGK model. This model was further simplified by Marle \cite{marle1969} and Anderson-Witting \cite{anderson1974relativistic} to calculate the transport coefficients. In the non-relativistic limit, Marle’s formulation leads to the same transport coefficient as the BGK model but fails in the relativistic limit. On the other hand, the Anderson-Witting model, also known as the relaxation-time approximation (RTA), is better suited in the relativistic limit. The RTA has been employed extensively in several areas of physics with considerable success and has been widely employed in the formulation of relativistic dissipative hydrodynamics \cite{Denicol:2010xn, Denicol:2012cn, Jaiswal:2014isa, Jaiswal:2015mxa, Florkowski:2012as, Florkowski:2013lza, Florkowski:2013lya, Denicol:2014mca, Florkowski:2014sfa, Florkowski:2014bba, Tinti:2015xwa, Czajka:2017wdo, Kurian:2018dbn, Chattopadhyay:2018apf, Chattopadhyay:2021ive, Jaiswal:2021uvv, Liyanage:2022nua}.

The RTA Boltzmann equation has provided remarkable insights into the causal theory of relativistic hydrodynamics as well as a simple yet meaningful picture of the collision mechanism in a non-equilibrium system. On the other hand, the BGK collision term ensures conservation of net particle four-current by construction, and is the precursor to RTA. Moreover, collision kernel in RTA involves only one timescale, the relaxation time, which governs all dissipation in the system. As a consequence, if the coefficient of shear viscosity is specified, the coefficient of bulk viscosity gets fixed automatically via the relaxation time. While the RTA has been employed extensively, a consistent formulation of relativistic dissipative hydrodynamics with the BGK collision term is relatively less explored. This may be attributed to the fact that the BGK collision kernel is ill defined for  relativistic systems without a conserved net particle four-current. This has limited the use of the BGK collision kernel to the studies related to flow of particle number and/or charge \cite{Carrington:2003je, Schenke:2006xu, Mandal:2013jla, Jiang:2016dkf, Han:2017nfz, Kumar:2017bja, Khan:2020rdw, Formanek:2021blc, Khan:2022orx, Khan:2022apd, Shaikh:2022sky}. 

In this article, we take the first step towards formulating a consistent framework of relativistic dissipative hydrodynamics using the BGK collision kernel. Furthermore, we propose a modified BGK collisions kernel (MBGK), which is well defined even in the absence of conserved particle four-current and is better suited for the formulation of the relativistic dissipative hydrodynamics. We find that there exists a free scalar parameter arising from the freedom of matching condition. This affects the scalar dissipation in the system, i.e., the coefficient of bulk viscosity. This feature allows us to specify the coefficient of bulk viscosity, independently from shear viscosity, as opposed to RTA. We study the effect on bulk viscosity in several different scenarios.

\section{Relativistic dissipative hydrodynamics}
\label{sec:RH}

The conserved net particle four-current, $N^\mu$, and the energy-momentum tensor, $T^{\mu\nu}$, of a system can be expressed in terms of the single particle phase-space distribution function and the hydrodynamic variables as \cite{DeGroot:1980dk},
\begin{align}
    N^\mu &= \! \int \mathrm{dP}\, p^\mu \left( f - \Bar{f} \right)= n\, u^\mu + n^\mu, \label{N^m_def}\\
    T^{\mu\nu} &=\! \int\! \mathrm{dP}\, p^\mu p^\nu \!\left( f + \Bar{f} \right)\! \nn\\&= \epsilon\, u^\mu u^\nu \!- \left(\! P_0 \!+\! \delta P \right)\! \Delta^{\mu\nu} \!+ \pi^{\mu\nu}\!, \label{T^mn_def}
\end{align}
where the Lorentz invariant momentum integral measure is defined as $\mathrm{dP} = g\, d^3p/\left[(2 \pi)^3 E\right]$ with $g$ being the degeneracy factor and $E = \sqrt{|{\bf p}|^2 + m^2}$ being the on-shell energy of the constituent particle of the medium with three-momentum ${\bf p}$ and mass $m$. Here $f \equiv f (x, p)$ and $\Bar{f} \equiv \Bar{f} (x, p)$ are the phase-space distribution functions for particles and anti-particles, respectively. In the above equations, $n$ is the net particle number density, $\epsilon$ is the energy density, $P_0$ is the equilibrium pressure, $n^\mu$ is the particle diffusion four-current, $\delta P$ is the correction to the isotropic pressure, and $\pi^{\mu\nu}$ is the shear stress tensor. We note that the fluid four-velocity $u^\mu$ has been defined in the Landau frame, $u_\mu T^{\mu\nu} = \epsilon u^\nu$. We also define $\Delta^{\mu\nu} \equiv g^{\mu\nu} - u^\mu u^\nu$ as the projection operator orthogonal to $u^\mu$. In this article, we will be working in a flat space-time with metric tensor defined as, $g^{\mu\nu} = {\rm diag}(1, -1, -1, -1)$.

Hydrodynamic equations are essentially the equations for conservation of net particle four current, $\partial_\mu N^\mu=0$, and energy-momentum tensor, $\partial_\mu T^{\mu\nu}=0$. Using the expressions of $N^\mu$ and $T^{\mu\nu}$ from Eqs.~\eqref{N^m_def} and \eqref{T^mn_def}, the hydrodynamic equations can be obtained as,
\begin{align}
    \Dot{n} + n \theta + \partial_\mu n^\mu &= 0 \label{heq1}\\
    \Dot{\epsilon} + \left( \epsilon + P_0 + \delta P \right) \theta - \pi_{\mu\nu} \sigma^{\mu\nu}& = 0 \label{heq2}\\
    \left( \epsilon + P_0 + \delta P \right) \Dot{u}^\alpha - \nabla^\alpha \left( P_0 + \delta P \right)\nn\\ + \Delta^\alpha_\nu \partial_\mu \pi^{\mu\nu} &= 0 \label{heq3}
\end{align}
where we use the standard notation, $\Dot{A} \equiv u_\mu \partial^\mu A$ for the co-moving derivatives, $\nabla^\alpha \equiv \Delta^{\alpha\beta} \partial_\beta$ for the space-like derivatives, $\theta = \partial_\mu u^\mu$ for the expansion scalar, and $\sigma^{\mu\nu} = \frac{1}{2} \left(\nabla^\mu u^\nu + \nabla^\nu u^\mu\right) - \frac{1}{3} \Delta^{\mu\nu} \theta$ for the velocity stress-tensor.

To express the conserved net particle four-current and the energy-momentum tensor in terms of hydrodynamic variables in Eqs.~\eqref{N^m_def} and \eqref{T^mn_def}, we chose Landau frame to define the fluid four-velocity. Additionally, the net-number density and energy density of a non-equilibrium system needs to be defined using the so called matching conditions. We relate these non-equilibrium quantities with their equilibrium values as
\begin{align}\label{n,e def}
    n = n_0 + \delta n, \qquad \epsilon = \epsilon_0 + \delta \epsilon, 
\end{align}
where $n_0$ and $\epsilon_0$ are the equilibrium net-number density and the energy density, respectively, and, $\delta n$, $\delta\epsilon$ are the corresponding non-equilibrium corrections. For a system which is out-of-equilibrium, the distribution function can be written as $f=f_0+\delta f$, where $f_0$ is the equilibrium distribution function and $\delta f$ is the non-equilibrium correction. In the present work, we consider the equilibrium distribution function to be of the classical Maxwell-Juttner form, $f_0=\exp(-\beta\, u\cdot p+\alpha)$, where $\beta\equiv 1/T$ is the inverse temperature, $\alpha\equiv\mu/T$ is the ratio of chemical potential to temperature and $u\cdot p\equiv u_\mu p^\mu$. The equilibrium distribution for anti-particles is also taken to be of the Maxwell-Juttner form with $\alpha\to-\alpha$.

We can now express the equilibrium hydrodynamic quantities in terms of the equilibrium distribution function as,
\begin{align}
    n_0 &= \int \mathrm{dP} \left(u\cdot p\right) \left( f_0 - \bar{f}_0\right) \label{neq}\\
    \epsilon_0 &= \int \mathrm{dP} \left(u\cdot p\right)^2 \left( f_0 + \bar{f}_0\right) \label{eeq}\\
    P_0 &= -\frac{1}{3}\Delta_{\mu\nu}\int \mathrm{dP}\, p^\mu p^\nu \left( f_0 + \bar{f}_0\right). \label{peq}
\end{align}
Similarly, the non-equilibrium quantities can be expressed as
\begin{align}
    \delta n &= \int \mathrm{dP} \left(u\cdot p\right) \left( \delta f - \delta \Bar{f} \right) \label{deln}\\
    \delta\epsilon &= \int \mathrm{dP} \left(u\cdot p\right)^2 \left( \delta f + \delta \Bar{f} \right) \label{dele}\\
    \delta P &= - \frac{1}{3} \Delta_{\alpha\beta} \int \mathrm{dP}\, p^\alpha p^\beta \left( \delta f + \delta \Bar{f} \right), \label{delp}\\
    n^\mu &= \Delta^\mu_\alpha \int \mathrm{dP}\, p^\alpha \left( \delta f - \delta \Bar{f} \right), \label{nmu}\\
    \pi^{\mu\nu} &= \Delta^{\mu\nu}_{\alpha\beta} \int \mathrm{dP}\, p^\alpha p^\beta \left( \delta f + \delta \Bar{f} \right),\label{pimunu}
\end{align}
where $\Delta^{\mu\nu}_{\alpha\beta}\equiv \frac{1}{2}(\Delta^{\mu}_{\alpha}\Delta^{\nu}_{\beta} + \Delta^{\mu}_{\beta}\Delta^{\nu}_{\alpha}) - \frac{1}{3}\Delta^{\mu\nu}\Delta_{\alpha\beta}$ is a traceless symmetric projection operator orthogonal to $u^\mu$ as well as $\Delta^{\mu\nu}$. In order to calculate the non-equilibrium quantities defined in Eqs.~\eqref{deln}-\eqref{pimunu}, we require the out-of-equilibrium correction to the distribution function i.e., $\delta f$ and $\delta\bar{f}$. For this purpose in the following, we consider the Boltzmann equation with BGK collision kernel, which describes the spacetime evolution of the distribution function.

\section{The Boltzmann equation and conservation laws}
\label{sec:BGK}

The covariant Boltzmann equation, in absence of any force term or mean-field interaction term, is given by,
\begin{align}
    p^\mu \partial_\mu f = C[f, \Bar{f}], \qquad p^\mu \partial_\mu \Bar{f} = \Bar{C}[f, \Bar{f}], \label{Beq}
\end{align}
for a single species of particles and its antiparticles. In the above equation, $C[f, \Bar{f}]$ and $\Bar{C}[f, \Bar{f}]$ are the collision kernels that contain the microscopic information of the scattering processes. For the formulation of relativistic hydrodynamics from the kinetic theory of unpolarized particles, the collision kernel of the Boltzmann equation must satisfy certain properties. Firstly, the collision kernel must vanish for a system in equilibrium, i.e., $C[f_0, \Bar{f}_0] = \Bar{C}[f_0, \Bar{f}_0] = 0$. Further, in order to satisfy the fundamental conservation equations in the microscopic interactions, the zeroth and the first moments of the collision kernel must vanish, i.e., $\int \mathrm{dP}\, C = 0$ and $\int \mathrm{dP}\, p^\mu\, C = 0$. Vanishing of the zeroth moment and the first moment of the collision kernel follows from the net particle four-current conservation and the energy-momentum conservation, respectively.

In the present work, we consider the BGK collision kernel which has the advantage that the particle four-current is conserved by construction. The relativistic Boltzmann equation with BGK collision kernel for particles can be written as \cite{Khan:2020rdw, Formanek:2021blc, Khan:2022orx, Khan:2022apd, Shaikh:2022sky},
\begin{align}
    p^\mu \partial_\mu f = - \frac{(u\cdot p)}{\tau_{\mathrm{R}}} \left( f - \frac{n}{n_0} f_0 \right),  \label{Beq_BGK}
\end{align}
and similarly for anti-particles with $f\to\bar{f}$ and $f_0\to\bar{f}_0$. Here, $\tau_{\mathrm{R}}$ is a relaxation time like parameter\footnote{A more conventional notation is the collision frequency which is defined as $\nu = 1/\tau_{\mathrm{R}}$.} which we assume to be the same for particles and anti-particles. It is easy to verify that the conservation of net particle four-current, defined in Eq.~\eqref{N^m_def}, follows from the zeroth moment of the above equations. The first moment of the above equations should lead to the conservation of the energy-momentum tensor, defined in Eq.~\eqref{T^mn_def}. However, we find that the first moment of the Boltzmann equation, Eq.~\eqref{Beq_BGK}, leads to,
\begin{align}
    \partial_\mu T^{\mu\nu} &= - \frac{1}{\tau_{\mathrm{R}}} \left( \epsilon - \frac{n}{n_0} \epsilon_0\right)u^\nu, \label{BGK_dT^mn}
\end{align}
which does not vanish automatically. 

In order to have energy-momentum conservation fulfilled by the Boltzmann equation with the BGK collision kernel, Eq.~\eqref{Beq_BGK}, we require that
\begin{align}
    \epsilon n_{0} = \epsilon_{0} n\,, \label{BGK_MC}
\end{align}
which we identify as one matching condition. Using Eq.~\eqref{n,e def}, the above constraint relation can be expressed in terms of  the out-of-equilibrium contribution to energy density and net number density as
\begin{equation}
   n_0 \delta\epsilon  = \epsilon_0  \delta n\,.\label{BGK_MC_1}
\end{equation}
Note that, in the present work, the fluid four-velocity has been defined using Landau frame condition. In order to derive a consistent hydrodynamic description from kinetic theory, one further requires two independent matching conditions to define two hydrodynamic field variables namely, the temperature, $T(x)$ and the chemical potential, $\mu(x)$~\cite{Kovtun:2019hdm}. This requirement essentially stems from the absence of a first-principle microscopic definition for nonequilibrium temperature and chemical potential. In the case of traditional RTA Boltzmann equation, $p^\mu\partial_\mu f=-\frac{(u\cdot p)}{\tau_R}(f-f_0)$, the two constraints necessary for the energy-momentum and net particle four-current conservation in the Landau frame are $\epsilon=\epsilon_0$ and $n=n_0$. Moreover, these matching conditions are also sufficient to define $T(x)$ and $\mu(x)$ in the RTA case. However, in the BGK approach, the net particle four-current conservation is automatic and does not lead to any constraint. Thus the only necessary condition obtained by demanding the  energy-momentum conservation  is not sufficient to specify both $T(x)$ and $\mu(x)$. In fact, Eq.~\eqref{BGK_MC} specifies one relationship among $T(x)$ and $\mu(x)$. However, the second independent matching condition is not determined by fundamental conservation laws and remains unconstrained within this framework. In this regard, it is interesting to note that the RTA framework is recovered if this second condition is fixed as $n=n_0$, or equivalently, $\epsilon=\epsilon_0$, demonstrating that the RTA prescription is a special case of the BGK framework. We shall see later that the scalar freedom due to the unconstrained second matching condition affects the coefficient of bulk viscosity, which is the transport coefficient corresponding to scalar dissipation in the system.

So far, we have considered the case of non vanishing chemical potential. On the other hand, the equilibrium net number density, defined in Eq.~\eqref{neq}, vanishes in the limit of zero chemical potential. This implies that the BGK collision term in Eq.~\eqref{Beq_BGK} is ill defined in this limit. However it is important to note that the limit of vanishing chemical potential is relevant for ultra-relativistic heavy-ion collisions. Therefore, it is desirable to modify the BGK collision kernel in order to extend its regime of applicability. At this juncture, we are well equipped to propose a modification to BGK collision kernel that is well-defined for all values of chemical potential. To this end, we rewrite the condition necessary for energy-momentum conservation from BGK collision kernel, Eq.~\eqref{BGK_MC}, in the form
\begin{align}
    \frac{n}{n_0} = \frac{\epsilon}{\epsilon_0}. \label{BGK_MC_R}
\end{align}
Substituting the above equation in Eq.~\eqref{Beq_BGK}, we obtain Boltzmann equation for particles with a modified BGK (MBGK) collision kernel,
\begin{align}
    p^\mu \partial_\mu f = - \frac{(u\cdot p)}{\tau_{\mathrm{R}}} \left( f - \frac{\epsilon}{\epsilon_0} f_0 \right),
    \label{MBGK_Beq}
\end{align}
and similarly for anti-particles with $f\to\bar{f}$ and $f_0\to\bar{f}_0$. The advantage of the above modification is that the collision kernel conserves energy-momentum by construction and is applicable to systems even without any conserved four-current, i.e., in the limit of vanishing chemical potential. Additionally, MBGK is uniquely defined even for a system with multiple conserved charges, as opposed to the BGK collision kernel. In the case of finite chemical potential, the matching condition, Eq.~\eqref{BGK_MC}, ensures net particle four-current conservation. It is important to note that BGK and MBGK are completely equivalent for the purpose of the derivation of hydrodynamic equations at finite chemical potential. Therefore, similar to the BGK framework, we require a second matching condition, along with the matching condition given in Eq.~\eqref{BGK_MC_1}, to formulate dissipative hydrodynamics using MBGK description. In the following, we consider the MBGK Boltzmann equation, Eq.~\eqref{MBGK_Beq}, to obtain non-equilibrium correction to the distribution function.

\section{Non-equilibrium correction to the distribution function} 
\label{sec:NECf}

In order to obtain the non-equilibrium correction to the distribution function, we use Eq.~\eqref{n,e def} to rewrite the MBGK Boltzmann equation, Eq.~\eqref{MBGK_Beq}, as
\begin{align}
    p^\mu \partial_\mu f = - \frac{(u\cdot p)}{\tau_{\mathrm{R}}} \left( \delta f - \frac{\delta \epsilon}{\epsilon_0} f_0 \right), \label{Beq2}
\end{align}
and similarly for anti-particles. The next step is to solve the 
above equation, order-by-order in gradients. In this work, we 
intend to obtain the non-equilibrium correction to the 
distribution function up to first-order in derivative, which we 
represent by $\delta f_1$. However, obtaining the expressions 
for $\delta f_1$ from Eq.~\eqref{Beq2} is not straightforward 
because it contains $\delta\epsilon$ which is defined in 
Eq.~\eqref{dele} as an integral over $\delta f$. Therefore, to solve for $\delta f_1$, we examine each term individually. At first,
we use Eqs.~\eqref{neq}-\eqref{peq} to substitute equilibrium hydrodynamic quantities in the conservation Eqs.~\eqref{heq1}-\eqref{heq3}, and obtain, up to first-order in gradients,
\begin{align}
    \Dot{\alpha} &=\chi_a\, \theta, \,~
    \Dot{\beta} =\chi_b\, \theta, \,~
    \nabla_\mu \beta = \frac{n_0}{\epsilon_0\!+\!P_0}\! \nabla_\mu \alpha \!- \beta \dot{u}_\mu,  \label{d_ab}
\end{align}
where,
\begin{equation}
\begin{aligned}
    \chi_a &= \frac{I_{20}^- (\epsilon_0 \!+\! P_0) \!- I_{30}^+ n_0}{I_{30}^+ I_{10}^+ - I_{20}^- I_{20}^-}, \\
    \chi_b &= \frac{I_{10}^+ (\epsilon_0 \!+\! P_0) \!- I_{20}^- n_0}{I_{30}^+ I_{10}^+ -I_{20}^- I_{20}^-}\label{chi_ab}~.
\end{aligned}
\end{equation}
The thermodynamic integrals are given by,
\begin{align}
    I^{\pm}_{nq} =\frac{(-1)^q}{(2q+1)!!}\int \mathrm{dP} &(u\cdot p)^{n-2q} \left(\Delta_{\alpha\beta} p^\alpha p^\beta \right)^q \nonumber\\
    &\times \left(f_0 \pm \Bar{f_0}\right)~, \label{I_nq}
\end{align}
and we identify $n_0 = I^-_{10}$, $\epsilon_0 = I^+_{20}$ and $P_0 = I^+_{21}$.
Replacing $\Dot{\alpha}, \Dot{\beta} ,\nabla_\mu \beta$ from Eq.~\eqref{d_ab}, the left-hand side of Eq.~\eqref{Beq2} becomes,
\begin{align}
    p^\mu \partial_\mu f_0 &=\left( A_{\Pi} \theta + A_{n} p^\mu \nabla_\mu \alpha + A_{\pi} p^\mu p^\nu \sigma_{\mu\nu} \right) f_0, \label{Beq_LHS_decomp}
\end{align}
where,
\begin{align}
    A_{\Pi} &= - \left[ \left(u \cdot p \right)^2 \left( \chi_b - \frac{\beta}{3} \right) - \left(u \cdot p \right) \chi_a + \frac{\beta m^2}{3} \right], \label{A_P} \\
    A_{n} &= 1 - \frac{n_0 \left(u \cdot p \right)}{\left( \epsilon_0 + P_0 \right)},\qquad
    A_{\pi} = -\, \beta . \label{A_n,p}
\end{align}
Similar procedure can be followed for anti-particles.
We assume $\delta f_1$ to have the same form as in Eq.~\eqref{Beq_LHS_decomp},
\begin{align}
    \delta f_1 &= \tau_{\rm R} \left( B_{\Pi} \theta + B_{n} p^\mu \nabla_\mu \alpha + B_{\pi} p^\mu p^\nu \sigma_{\mu\nu} \right) f_0, \label{del f_1 decomp}
\end{align}
and similarly for anti-particles. In the above expression, the coefficients $B_{\Pi}$, $B_{n}$ and $B_{\pi}$ needs to be determined using Eq.~\eqref{Beq2}, up to first order in derivatives. At this point, we note that some discussion is due on the choice of the structure of $\delta f_1$. In general, we may write $\delta f_1 = \phi_1 f_0$, with,
\begin{align}
    \phi_1 = \sum_{\ell = 0}^\infty B_\ell\, p^{\langle \mu_1} \cdots p^{\mu_\ell\rangle} G_{\mu_1\cdots\mu_\ell}, \label{delf-expand}
\end{align}
where $G_{\mu_1\cdots\mu_\ell}$ is some $\ell$-rank gradient \cite{DeGroot:1980dk}, $A^{\langle\mu_1} \cdots A^{\mu_\ell\rangle} = \Delta^{\mu_1\cdots \mu_\ell}_{\nu_1\cdots \nu_\ell} A^{\nu_1} \cdots A^{\nu_\ell}$, and $B_\ell$ is the coefficient of the $\ell$-th rank gradient, which can be function of particle momenta. Substituting Eq.~\eqref{delf-expand} in Eq.~\eqref{Beq2}, and comparing both sides of the Boltzmann equation, one may conclude $B_\ell = 0$ for $\ell\geq3$, where we also have to recall the scalar nature of $\delta\epsilon$.

It should be noted here that the modified collision kernel in Eq.~\eqref{Beq2} possess a term proportional to $\delta \epsilon$ which also needs to be determined up to first order in gradient. For that purpose we substitute the above decomposition of $\delta f_1$ and the corresponding decomposition for anti-particles in the definition of $\delta \epsilon$  given in Eq.~\eqref{dele}. It can be shown that the contributions from both the vector and tensor components drop out due to their orthogonality properties. Only the contribution from the scalar gradient survives and is given by
\begin{align}
    \delta \epsilon = \tau_{\rm R} \int \mathrm{dP} \left(u \cdot p\right)^2 \Big( B_{\Pi} f_0 + \Bar{B}_{\Pi} \Bar{f}_0 \Big)\, \theta.\label{del_e}
\end{align}
Using Eqs.~\eqref{Beq_LHS_decomp}, \eqref{del f_1 decomp} and \eqref{del_e} into Eq.~\eqref{Beq2} and comparing both sides, we get
\begin{align}
    &- \frac{A_{\Pi}}{(u\cdot p)} = B_{\Pi} -\frac{1}{\epsilon_0} \!\int\! \mathrm{dP}\, \left( u\cdot p\right)^2 \Big( B_{\Pi} f_0 + \bar{B}_{\Pi} \bar{f}_0 \Big), \label{AP-BP_rel}\\
    &B_{n}=-\frac{A_{n}}{(u\cdot p)}\,, \qquad B_{\pi}=-\frac{A_{\pi}}{(u\cdot p)}. \label{An,p-Bn,p_rel}
\end{align}
Another set of equations in terms of $\bar{A}_\Pi$, $\bar{A}_n$ and  $\bar{A}_\pi$ can be obtained by considering MBGK equation, analogous to Eq.~\eqref{Beq2}, for anti-particles. Here, coefficients $B_{n}$, $\Bar{B}_{n}$, $B_{\pi}$ and $\Bar{B}_{\pi}$ are easily determined but $B_{\Pi}$ and $\Bar{B}_{\Pi}$ require further investigation.

To obtain their expressions, we consider $B_{\Pi}$ to be of the general form, $B_{\Pi} = \sum_{k=-\infty}^{+\infty} b_k \left(u \cdot p\right)^k$ and $\Bar{B}_{\Pi} = \sum_{k=-\infty}^{+\infty} \Bar{b}_k \left(u \cdot p\right)^k$. Substituting these in Eq.~\eqref{AP-BP_rel} and its corresponding equation for anti-particles, we can conclude that the only non-zero $b_k$ and $\bar{b}_k$ are the ones with $k=-1,0,1$. We obtain
\begin{align}
    B_{\Pi} = \sum_{k=-1}^1 b_k \left(u \cdot p\right)^k ,
    \qquad
    \Bar{B}_{\Pi} = \sum_{k=-1}^1 \Bar{b}_k \left(u \cdot p\right)^k. \label{B_1-bB_1}
\end{align}
Substituting Eqs.~\eqref{A_P} and \eqref{B_1-bB_1} in Eq.~\eqref{AP-BP_rel}, we find
\begin{align}
    &b_1 = \bar{b}_1 = \chi_b - \frac{\beta}{3} \qquad \mathrm{and,}\qquad b_{-1} = \bar{b}_{-1} = \frac{m^2 \beta}{3}, \label{b_1,-1}
\end{align}
where we have also used the relation analogous to Eq.~\eqref{AP-BP_rel} for anti-particles. On the other hand, for $b_0$ and $\Bar{b}_0$, we find two coupled equations which are identical and leads to the relation
\begin{align}
    \bar{b}_0 = b_0 + 2\, \chi_a. \label{b0-ab0_rel}
\end{align}
Hence, we see that a unique solution for $b_0$ and $\bar{b}_0$ can not be obtained but they are constrained by the above relation. We need to provide one more condition, which we recognize as the second matching condition, to fix $b_0$ and $\bar{b}_0$ separately. 

Nevertheless, at this stage, we can determine $\delta f_1$ and $\delta \bar{f}_1$ up to a free parameter, $b_0$, by using Eqs.~\eqref{AP-BP_rel}-\eqref{b0-ab0_rel} into Eq.~\eqref{del f_1 decomp}, and similarly for anti-particles. We obtain,
\begin{align}
   \frac{\delta f_1}{\tau_{\mathrm{R}}\, f_0 } =&\bigg[ \left\{ \frac{m^2 \beta}{3 \left(u\cdot p\right)} + b_0 + \left( u\cdot p\right) \left(\chi_b - \frac{\beta}{3}\right) \right\} \theta \nonumber\\
    &- \left\{ \frac{1}{\left(u\cdot p\right)} - \frac{n_0}{\left( \epsilon_0 + P_0 \right)} \right\} p^\mu \left(\nabla_\mu \alpha \right)\nonumber\\&+\frac{\beta p^\mu p^\mu \sigma_{\mu\nu}}{\left(u\cdot p\right)} \bigg], \label{delf_1}\\
    \frac{\delta \bar{f}_1}{\tau_{\mathrm{R}}\, \Bar{f}_0} =&\bigg[\! \left\{\! \frac{m^2 \beta}{3 \left(u\!\cdot\!p\right)} + b_0 + 2\, \chi_a + \!\left( u\!\cdot\!p\right)\!\! \left(\!\chi_b - \frac{\beta}{3}\!\right)\! \right\}\! \theta \nonumber\\
    &+ \left\{ \frac{1}{\left(u\cdot p\right)} + \frac{n_0}{\left( \epsilon_0 + P_0 \right)} \right\} p^\mu \left(\nabla_\mu \alpha \right) \nonumber\\&+\frac{\beta p^\mu p^\mu \sigma_{\mu\nu}}{\left(u\cdot p\right)} \bigg]. \label{delaf_1}
\end{align}
Note that for vanishing chemical potential, we have $\alpha=\chi_a=0$. In this case, Eqs.~\eqref{delf_1} and \eqref{delaf_1} coincide to give
\begin{align}
   \frac{\delta f_1}{ \tau_{\mathrm{R}}f_0}\Big\vert_{\mu=0} = \beta\bigg[& \left\{\! \frac{m^2}{3 \left(u\!\cdot\! p\right)} + \frac{b_0}{\beta} + \!\left( u\!\cdot\! p\right)\! \left(c_s^2 \!-\! \frac{1}{3}\right) \right\}\! \theta \nonumber\\
    &+ \frac{p^\mu p^\mu \sigma_{\mu\nu}}{\left(u\!\cdot\! p\right)} \bigg], \label{delf_1_mu0}
\end{align}
where we have used $\chi_b = \beta c_s^2$, with $c_s^2$ being the squared of the speed of sound, given by,
\begin{align}
    c_s^2 = \frac{\left(\epsilon_0 + P_0 \right)}{3 \epsilon_0 + \left( 3 + z^2 \right) P_0}. \label{MBGK_cs2}
\end{align}
Here $z\equiv m/T$ is the ratio of particle mass to temperature.

\section{First order dissipative hydrodynamics}
\label{sec:RH_FODH}

The first-order correction to the phase-space distribution functions of the particles and anti-particles at finite $\mu$ are given by Eqs.~\eqref{delf_1} and \eqref{delaf_1}. Substituting them in Eqs.~\eqref{deln}-\eqref{pimunu}, we obtain the first-order expressions for non-equilibrium hydrodynamic quantities as
\begin{align}
    \delta n =&\ \nu \theta, \qquad \delta \epsilon = e \theta, \qquad \delta P = \rho \theta, \nonumber\\ 
    n^\mu =&\ \kappa \nabla^\mu \alpha, \qquad \pi^{\mu\nu} = 2 \eta \sigma^{\mu\nu}, \label{RH_NS}
\end{align}
where,
\begin{align}
    &\nu = \tau_{\mathrm{R}} \left( \chi_a + b_0 \right) n_0,
    \quad
    e= \tau_{\mathrm{R}} \left( \chi_a + b_0 \right) \epsilon_0, \label{RH_FODC_nu/ve} \\
    &\rho= \tau_{\mathrm{R}} \left[ (\chi_a + b_0) P_0 + \chi_b\frac{(\epsilon_0+P_0)}{\beta} - \frac{5}{3}\beta I_{32}^+\right. \nonumber\\&\left.\quad\quad\quad\quad - \frac{\chi_a n_0}{\beta} \right], \label{RH_FODC_zeta} \\
    &\kappa = \tau_{\mathrm{R}} \left[I_{11}^+-\frac{n_0^2}{\beta(\epsilon_0+P_0)} \right],
    \quad
    \eta = \tau_{\mathrm{R}}\, \beta\, I_{32}^+. \label{RH_FODC_kappa/eta}
\end{align}
Note that the parameter $b_0$ appears in the expressions of $\nu$, $e$ and $\rho$. Of these, $\nu$ and $e$ vanishes for $b_0=-\chi_a$ which corresponds to the Landau matching condition and RTA collision kernel. Conductivity $\kappa$ and the coefficient of shear viscosity $\eta$ does not contain the parameter $b_0$, and the expressions for these two transport coefficients, given in Eq.~\eqref{RH_FODC_kappa/eta}, matches with those derived using RTA collision kernel \cite{Jaiswal:2015mxa}. Next, we analyze entropy production in the MBGK setup in order to identify dissipative transport coefficients in Eqs.~\eqref{RH_FODC_nu/ve}-\eqref{RH_FODC_kappa/eta}.

To study entropy production, we start from the kinetic theory definition of entropy four-current, given by the Boltzmann's H-theorem, for a classical system
\begin{align}
    S^\mu = - \int \mathrm{dP} p^\mu \Big[ f \left(\ln{f} -1 \right) + \Bar{f} \left(\ln{\Bar{f}} - 1\right) \Big]. \label{H^mu_KTdef}
\end{align}
The entropy production is determined by taking four-divergence of the above equation,
\begin{align}
    \partial_\mu S^\mu\! =\! - \int \!\mathrm{dP} p^\mu \Big[ \left(\partial_\mu f \right) \ln{f} + \left(\partial_\mu \Bar{f} \right) \ln{\Bar{f}} \Big]. \label{del.H-KT}
\end{align}
Using the MBGK Boltzmann equation, i.e., Eq.~\eqref{Beq2}, and keeping terms till quadratic order in deviation-from-equilibrium, we obtain
\begin{align}
    \partial_\mu S^\mu = \frac{1}{\tau_{\rm R}} \!\int\! \mathrm{dP} \left(u\!\cdot\!p\right)& \!\left[ \!\left(\! \phi - \frac{\delta \epsilon}{\epsilon_0}\!\right)\! \phi f_0\right. \nonumber\\&\left. + \!\left(\bar{\phi} - \frac{\delta \epsilon}{\epsilon_0} \!\right)\! \bar{\phi} \Bar{f}_0 \right]\!. \label{del.H-MBGK}
\end{align}
where we have defined $\phi\equiv\delta f/f_0$ and $\bar{\phi}\equiv\delta\bar{f}/\bar{f}_0$.

Using Eqs.~\eqref{delf_1} and \eqref{delaf_1} in Eq.~\eqref{del.H-MBGK}, we obtain,
\begin{align}
    \partial_\mu S^\mu = - \beta\, \Pi\, \theta - n^\mu \nabla_\mu \alpha + \beta \pi^{\mu\nu} \sigma_{\mu\nu}, \label{del.H_final}
\end{align}
where,
\begin{align}
    \Pi &= \delta P - \frac{\chi_b}{\beta} \delta \epsilon + \frac{\chi_a}{\beta} \delta n. \label{H-Pi}
\end{align}
Note that the right-hand-side of Eq.~\eqref{del.H_final} represents entropy production due to dissipation in the system. Here the shear stress tensor $\pi^{\mu\nu}$ is the tensor dissipation, the particle diffusion four-current $n^\mu$ is the vector dissipation and $\Pi$ is the scalar dissipation, referred to as the bulk viscous pressure\footnote{We can further identify that, $\left( \frac{\partial P}{\partial \epsilon} \right)_n = \frac{\chi_b}{\beta}$ and, $\left( \frac{\partial P}{\partial n} \right)_\epsilon = - \frac{\chi_\alpha}{\beta}$.}. From Eq.~\eqref{H-Pi}, we observe that  $\delta P$, $\delta \epsilon$, and $\delta n$, all contribute to the bulk viscous pressure. Comparing with the Navier-Stokes relation
$\Pi = - \zeta\, \theta$, the coefficient of bulk viscosity becomes,
\begin{align}
    \zeta = - &\tau_{\rm R}\bigg[ \frac{\chi_b}{\beta} \left( \epsilon_0 + P_0\right) - \frac{5 \beta I_{32}^+}{3} -\frac{\chi_a n_0}{\beta} \nonumber\\
    &+ \frac{(\chi_a+b_0)}{\beta} \left(\, \beta P_0 - \chi_b \epsilon_0 + \chi_a n_0 \right) \bigg]. \label{RH_zeta}
\end{align}
Demanding that Eq.~\eqref{del.H_final} does not violate the second law of thermodynamics, i.e., $\partial_\mu S^\mu \geq 0$, leads to the following constraints ~\cite{Kovtun:2019hdm},
\begin{align}
    \zeta \geq 0,
    \quad
    \kappa \geq 0,
    \quad
    \eta \geq 0. \label{TC_positivity}
\end{align}
These three transport coefficients represent the three dissipative transport phenomena of the system related to the transport of momentum and charge. We see that out of the three transport coefficients, only $\zeta$ depends on the parameter $b_0$ and the second matching condition is necessary to uniquely determine $\zeta$. This is to be expected because the matching conditions are scalar conditions and should only affect the scalar dissipation in the system, i.e., bulk viscosity. In the following, we specify the second matching condition.

With the parameter, $b_0$ still not specified, the hydrodynamic equations obtained using the MBGK Boltzmann equation forms a class of hydrodynamic theories. A specific hydrodynamic theory is determined by a specific $b_0$ parameter. We can access different hydrodynamic theories by varying the $b_0$ parameter, which is solely controlled by the second matching condition. Thus, picking a specific second matching condition will fix $b_0$ and hence the hydrodynamic theory. To this end, we define a function $\mathcal{A}_r^\pm$ as \cite{Rocha:2021zcw, Biswas:2022cla},
\begin{align}
    \mathcal{A}_r^\pm = \int \mathrm{dP}\, (u\cdot p)^{r} \left(\delta f \pm \delta \Bar{f} \right). \label{RH_AMC_genMC}
\end{align}
The second matching condition then amounts to assigning a value for a given $\mathcal{A}_r^\pm$. For instance, the RTA matching conditions can be recovered by setting $\mathcal{A}_1^- = \mathcal{A}_2^+ = 0$. It is apparent that the choice of a second matching condition is vast, and determination of the full list of the allowed ones is a non-trivial task that goes beyond the scope of the present work. Presently, for the second matching condition, we shall restrict our analysis to a special set $\mathcal{A}_r^+=0$. These matching conditions ensures that the {\it homogeneous} part of $\delta f$ vanishes\footnote{It must be noted that this is not the only class of matching conditions that guarantee the zero value of the homogeneous part.} \cite{Hoult:2021gnb} and are also valid in the zero chemical potential limit. Using Eqs.~\eqref{delf_1} and \eqref{delaf_1} in our proposed matching condition $\mathcal{A}_r^+=0$, we obtain 
\begin{equation}
    \begin{aligned}
     b_0 = - \left(1/I_{r,0}^+\right) &\left[ \chi_b I_{r + 1,0}^+ - \beta I_{r+1,1}^+\right.\\ &\left.+ \chi_a \left(I^+_{r,0} - I^-_{r,0}\right) \right].
    \label{RH_AMC_b0}
    \end{aligned}
\end{equation}

We emphasize that the above equation along with Eq.~\eqref{BGK_MC} constitute the two matching conditions required to define temperature and chemical potential of the system. In order to compare with the usual Landau matching conditions, we express Eq.~\eqref{n,e def} using first order results obtained in Eqs.~\eqref{RH_NS} and \eqref{RH_FODC_nu/ve},
\begin{align}
n &= \left[1+\tau_{\mathrm{R}}\! \left( \chi_a + b_0 \right)\theta \right] n_0(T,\mu) , \label{match_n} \\
\epsilon &= \left[1+\tau_{\mathrm{R}}\! \left( \chi_a + b_0 \right)\theta \right] \epsilon_0(T,\mu) . \label{match_e}
\end{align}
The above conditions reduce to the usual Landau matching condition, $n=n_0$ and $\epsilon=\epsilon_0$, for the particular choice of $b_0=- \chi_a$. In the next section, we explore the effect of different $b_0$ on the coefficient of bulk viscosity.

\section{Results and discussions}
\label{sec:res_disc}

In this section, we study the effect of MBGK collision kernel on transport coefficients. In the previous Section, we found that the effect of MBGK collision kernel manifests in the parameter $b_0$ which affects only the scalar dissipation, namely bulk viscous pressure. On the other hand, the vector (net particle diffusion) and tensor (shear stress tensor) dissipation remain unaffected. Therefore, we study only the properties of bulk viscous coefficient in this section.


\begin{center}
    \begin{figure}[t]
        \begin{center}
            \includegraphics[width=\linewidth]{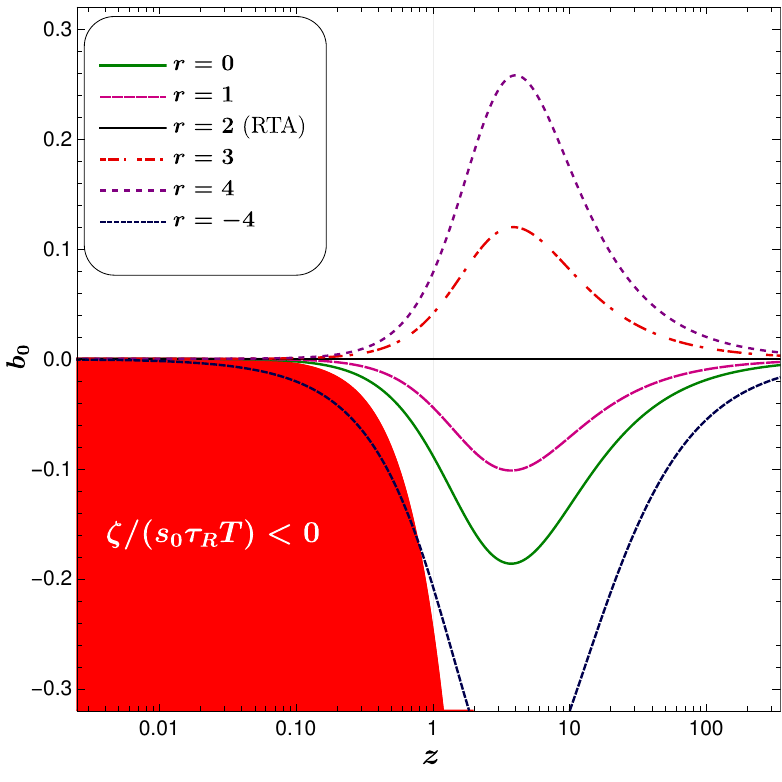}
            \vspace{-0.7cm}
            \caption{Dependence of the parameter $b_0$ on $z$ for different matching conditions. The red region corresponds to negative values of $\zeta$. The plot is for zero chemical potential.}
            \label{contour}
            \vspace{-0.2cm}
        \end{center}
    \end{figure}
\end{center}

Before we proceed to quantify the effect of varying the second matching condition on the coefficient of bulk viscosity, we must establish the allowed values for the parameter $b_0$. To this end, we note that the second law of thermodynamics demands that the coefficient of bulk viscosity must be positive, Eq.~\eqref{TC_positivity}. In Fig.~\ref{contour}, we plot $b_0$ vs $z$ for different values of $r$ required to define the second matching condition in Eq.~\eqref{RH_AMC_b0}, at zero chemical potential. The red region in Fig.~\ref{contour} corresponds to the part of $b_0$-$z$ plane where the coefficient of bulk viscosity becomes negative. Therefore all values of $r$ for which the curves for $b_0$ lies in the red zone are not physical and must be discarded. The boundary of the red region corresponds to the $\zeta = 0$ line and is 
\begin{align}
    b_0^l = - \chi_a + \left[ \frac{\chi_b \left(\epsilon_0 + P_0 \right) - \chi_a n_0 - \left(5/3\right) \beta^2 I_{32}^+ }{\, \chi_b \epsilon_0 - \chi_a n_0 - \beta P_0 } \right]. \label{b0->zeta=0}
\end{align}
Note that $b_0^l$ as given in Eq.~\eqref{b0->zeta=0} is defined in terms of the ratio of thermodynamic quantities only. In order to ensure $\zeta\geq0$, we can consider $b_0\geq b_0^l(\alpha,z)$ as a physical constraint on the choice of $b_0$.  While one can use different parametrizations for $b_0$, this condition is more rigorous and independent of the choice of parametrization. With the current choice, we find that $b_0$ parameter with non-negative values of $r$ respects the requirement of the second law of thermodynamics, Eq.~\eqref{TC_positivity}, whereas, large negative values of $r$, violate the constraint, $b_0 \geq b_0^l$ leading to negative values of $\zeta$. In Fig.~\ref{contour}, we see that the curve for $b_0$, which corresponds to $r=-4$, passes through the physically forbidden region. The black line with $r=2$ represents the $b_0$ for which the MBGK reduces to the RTA, where $b_0$ vanishes for all $z$ . 


\begin{center}
    \begin{figure}[t]
        \begin{center}
            \includegraphics[width=\linewidth]{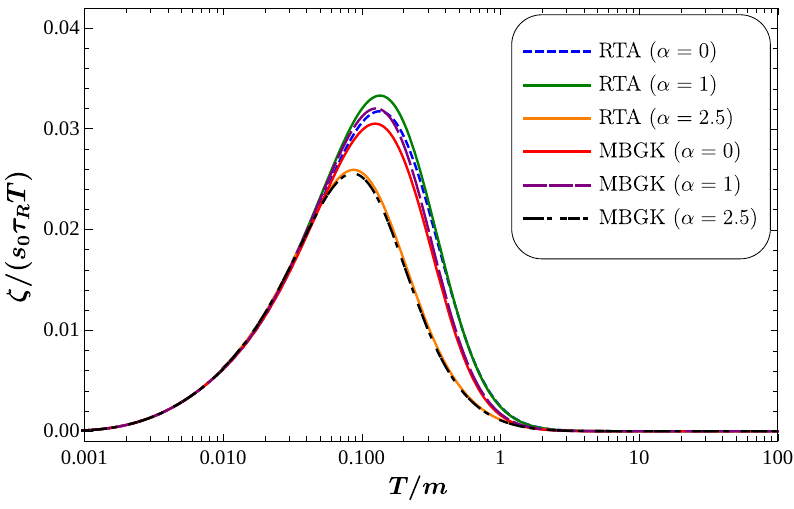}
            \vspace{-0.7cm}
            \caption{Dependence of $\zeta/\left( s_0 \tau_{\rm R} T\right)$ on the $T/m$ for various $\alpha = \mu/T$ values. The curves labelled RTA corresponds to $r=2$ and those labelled MBGK corresponds to $r=0$.}
            \label{zetabys}
            \vspace{-0.2cm}
        \end{center}
    \end{figure}
\end{center}

Having determined the allowed range of $r$ and equivalently, the allowed values of $b_0$, we will restrict ourselves to $b_0$ corresponding to $r\geq 0$ values. In Fig.~\ref{zetabys} we plot the dimensionless quantity $\zeta/\left(s_0 \tau_{\rm R}T\right)$ for MBGK with $r=0$, and RTA ($r=2$) against $T/m$ for different values of chemical potential, where $s_0\equiv(\epsilon_0+P_0-\mu\,n_0)/T$. We observe that $\zeta/\left(s_0 \tau_{\rm R}T\right)$ is a non-monotonous function of temperature, having a maximum for each $r$ for MBGK case, similar to the behavior known from RTA~\cite{Rocha:2021zcw, Biswas:2022cla, Dash:2021ibx}. We also note that the dependence of $\zeta/\left(s_0 \tau_{\rm R}T\right)$ on $\alpha$ is also non-monotonous, which can be realized by observing that not only the position of the peak for $\alpha=1$ is at higher $T/m$ values than for $\alpha=0$ and $\alpha=2.5$, but the peak value is also higher for $\alpha=1$ compared to $\alpha=0$ and $\alpha=2.5$. 

To better understand the effect of changing matching conditions on the behavior of the bulk viscosity for the MBGK collision kernel, we focus on the zero chemical potential limit. In this limit, we study the scaling behavior of the ratio of the coefficient of bulk viscosity to shear viscosity, $\zeta/\eta$, with conformality measure $1/3-c_s^2$. In Fig.~\ref{zeta/eta}, we plot the ratio $(\zeta/\eta)/(1/3-c_s^2)^2$ as a function of $z$ for different $r$ values. We observe that this ratio saturates in both small-$z$ and large-$z$ limits indicating a squared dependence of $\zeta/\eta$ on the conformality measure, characteristic to weakly coupled systems. We also observe that in the small-$z$ limit, this ratio saturates to different values whereas in the large-$z$ limit, they all converge. In order to better understand the behavior of $\zeta/\eta$ in these regimes, we separately analyze the small-$z$ and large-$z$ limits.

\subsection{Small-$z$ behaviour} 

\begin{center}
    \begin{figure}[t]
        \begin{center}
            \includegraphics[width=\linewidth]{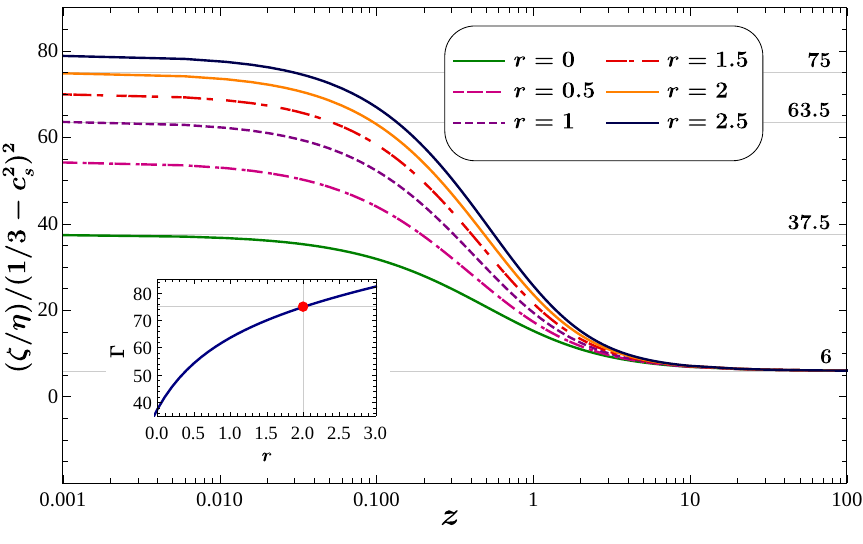}
            \vspace{-0.7cm}
            \caption{Variation of the dimensionless quantity $(\zeta/\eta)/(1/3 - c_s^2)^2$ with respect to $z$ for various matching conditions determined by $r$. Inset: Variation of the scaling coefficient $\Gamma$, defined in Eqs.~\eqref{zeta/eta_scaling} and \eqref{Gamma_MBGK}, with respect to parameter $r$. The red dot represents the RTA value of $\Gamma=75$.}
            \label{zeta/eta}
            \vspace{-0.2cm}
        \end{center}
    \end{figure}
\end{center}

The small-$z$ limit, i.e., $m/T\ll1$, is the ultra-relativistic limit where the mass of the particles can be ignored compared to the temperature of the system. At zero chemical potential, the small-$z$ limiting behavior of the conformality measure is given by $\left( \frac{1}{3} - c_s^2 \right) = \frac{z^2}{36} + \mathcal{O} \left(z^3\right)$. On the other hand, the small-$z$ behavior of the ratio $\zeta/\eta$ is found to be
\begin{align}
    \frac{\zeta}{\eta} = \Gamma(r) \left( \frac{1}{3} - c_s^2 \right)^2 + \mathcal{O} \left(z^5\right), \label{zeta/eta_scaling}
\end{align}
for all $r$. We find the $r$-dependence of the coefficient to be,
\begin{align}
    \Gamma(r) \equiv \lim_{z\to 0} \frac{\zeta/\eta}{\left( \frac{1}{3} - c_s^2 \right)^2} = \frac{15( r^2 + 23 r + 10)}{4 (r + 1)}, \label{Gamma_MBGK}
\end{align}
for $r \geq 0$. Thus, while the ratio $\zeta/\eta$ shows a $z^4$ dependence in the same small-$z$ limit, the coefficient $\Gamma$ depends on the matching condition through $b_0$, and equivalently $r$, as is evident from Eq.~\eqref{Gamma_MBGK}. In the inset of Fig.~\ref{zeta/eta}, we show the variation of the coefficient $\Gamma$ as a function of $r$. We observe that for $r=2$, we recover the RTA value, $\Gamma=75$, marked with a red dot. 


\subsection{Large-$z$ behaviour}

On the opposite end, i.e., at the large-$z$ limit where $m/T\gg1$, we have the non-relativistic limit. In this limit, the conformality measure is expanded in powers of $1/z$ and is given by, $\frac{1}{3} - c_s^2 = \frac{1}{3} - \frac{1}{z} +  \mathcal{O}\left(\frac{1}{z^2}\right)$ . The behaviour of the ratio $\zeta/\eta$ in the same limit is given by,
\begin{align}
    \frac{\zeta}{\eta} = \frac{2}{3} - \frac{3}{z} + \mathcal{O} \left(\frac{1}{z^2}\right), \label{zeta/eta_large-z}
\end{align}
for all $r$. Considering only the leading terms in this expansion, we find $(\zeta/\eta)/(1/3-c_s^2)^2=6$, as is evident from Fig.~\ref{zeta/eta}. Considering terms up to $1/z$ in the expansion, we get,
\begin{align}
    \frac{\zeta}{\eta} = 2\sqrt{3} \left( \frac{1}{3} - c_s^2 \right)^{3/2}, \label{scaled_zeta/eta_large-z}
\end{align}
which is independent of $r$ and hence the second matching condition. The above equation is the scaling relation we obtain in the non-relativistic limit. In this limit, the MBGK and RTA results coincide implying that the properties of the fluid are independent of the nature of collision with BGK collision kernel.

\begin{center}
    \begin{figure}[t]
        \begin{center}
            \includegraphics[width=\linewidth]{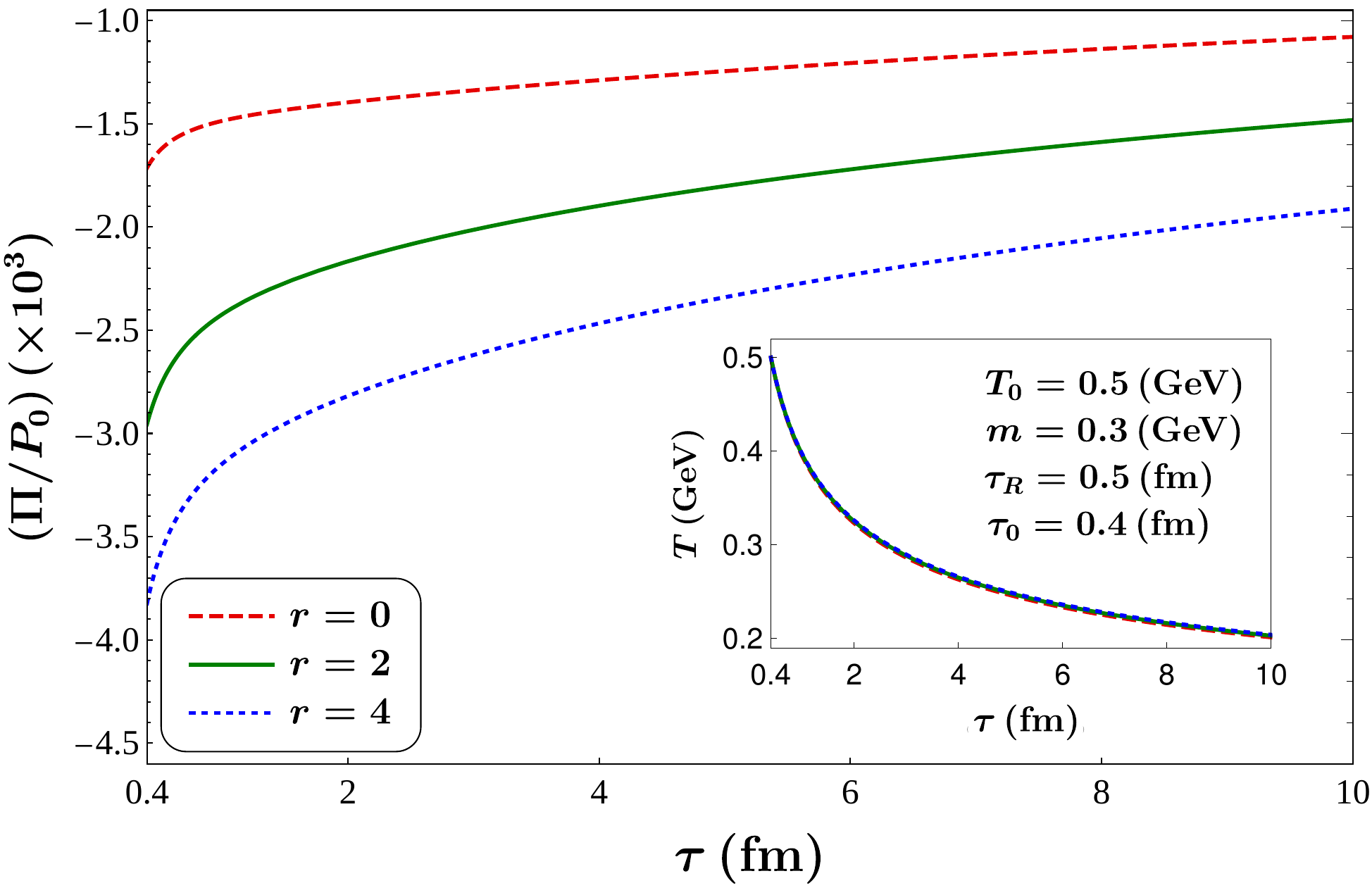}
            \vspace{-0.7cm}
            \caption{Proper time evolution of bulk viscous pressure, scaled by equilibrium pressure. Inset: Proper time evolution of temperature. Boost-invariant Bjorken expansion is considered to generate curves for different $r$ values.}
            \label{Pi/P0}
            \vspace{-0.2cm}
        \end{center}
    \end{figure}
\end{center}

\subsection{Bjorken expansion}

In order to study the effect of present hydrodynamic formulation on evolution of rapidly expanding medium, we consider the case of transversely homogeneous and purely longitudinal boost-invariant expansion, $v_z=z/t$~\cite{Bjorken:1982qr}. It is convenient to work in the Milne co-ordinate system, $(\tau,x,y,\eta_s)$, where $\tau=\sqrt{t^2-z^2}$ is the longitudinal proper time and $\eta_s = \tanh^{-1}\left(z/t\right)$ is the space time rapidity and the metric tensor is given by $g_{\mu\nu}=\left(1,-1,-1,-\tau^2\right)$. In this case, the fluid four velocity becomes $u^\mu=\left(1,0,0,0\right)$ and all functions of space and time depend only on $\tau$. For zero chemical potential, within MBGK framework, Eq.~\eqref{heq2}  can be written as,
\begin{align}
    \Dot{\epsilon}_0 + \delta \Dot{\epsilon} + \left( \epsilon_0 + P_0 \right) \theta + \left(\delta \epsilon + \delta P \right) \theta - \pi^{\mu\nu} \sigma_{\mu\nu} = 0, \label{heq2_full}
\end{align}
where, $\theta=1/\tau$, $\dot{\epsilon}=d\epsilon/d\tau$ and $\pi^{\mu\nu}\sigma_{\mu\nu}=4\eta/3\tau^2$ where $\eta$ is given in Eq.~\eqref{RH_FODC_kappa/eta}. The free parameter $b_0$ enters in the evolution equation through the scalar deviations $\delta \epsilon$ and $\delta P$ given in Eqs.~\eqref{RH_NS}-\eqref{RH_FODC_zeta}. Moreover, $\delta \dot{\epsilon}$ is given by,
\begin{align}
    \delta\dot{\epsilon}=\frac{\tau_\mathrm{R}}{\tau}\left[\left(\epsilon_0\dot{b}_0+b_0\dot{\epsilon}_0\right)-\frac{b_0\epsilon_0}{\tau}\right],
    \label{deleeq}
\end{align} 
where,
\begin{align}
    \dot{b}_0=&\dot{\beta}\left[b_0
    \left(\frac{1}{\beta}+
    \frac{I_{r+1,0}}{I_{r,0}}\right)-
    \beta\left\{-
    \left(\frac{c_s^2 I_{r+2,0}-
    I_{r+2,1}}{I_{r,0}}\right)\right.\right.\nonumber\\&\left.\left.+\frac{I_{r+1,0}}
    {I_{r,0}}\left(\frac{c_s^2 
    I_{4,0}-I_{4,1}}{I_{3,0}}\right)\right
    \}\right],
    \label{eqb0dot}
\end{align}
is obtained from our choice of $b_0$ in Eq.~\eqref{RH_AMC_b0} which, in case of $\alpha=0$ reduces to,
\begin{align}
    b_0 = - \left(\beta/I_{r,0}\right) \Big[ c_s^2 I_{r+1,0} - I_{r+1,1} \Big]\,.
    \label{b_0-def_mu=0}
\end{align} 
The energy evolution equation then takes the form, 
\begin{align}
    \dot{\epsilon}_0+&\frac{\tau_\mathrm{R}}{\tau}\left(\epsilon_0\dot{b}_0+b_0\dot{\epsilon}_0\right)-\frac{\tau_\mathrm{R}}{\tau^2}\left(b_0\epsilon_0+3\beta I_{32}\right)\nonumber\\&+\frac{\left(\epsilon_0+P_0\right)}{\tau}\left[1+\left(b_0+c_s^2\right)\frac{\tau_\mathrm{R}}{\tau}\right]=0~.
    \label{eneq_full}
\end{align}
Implementing the  $\dot{b}_0$  obtained in Eq.~\eqref{eqb0dot}, one can express Eq.~\eqref{eneq_full}   as
 \begin{align}
      \mathcal{G}_r\left(T(\tau),\tau\right)\frac{d T(\tau)}{d\tau} + T^2(\tau)\mathcal{H}_r(T(\tau),\tau)  = 0\,,
      \label{Teq_full}
 \end{align}
where, we have introduced the functions $\mathcal{G}_r$ and $\mathcal{H}_r$ for convenience. Their explicit expressions in terms of the thermodynamic integrals are given by
\begin{align}
    \mathcal{G}_r&=\left[ I_{3,0}- \frac{\tau_\mathrm{R}}{\tau}\left(\epsilon_0\mathcal{F}_r -b_0  I_{3,0}\right) \right]\,,
    \label{Gr}
\end{align}
with
\begin{align}
    \mathcal{F}_r&=\left[b_0\left(\frac{1}{\beta}+\frac{I_{r+1,0}}{I_{r,0}}\right)-\beta\left\{\frac{I_{r+1,0}}{I_{r,0}}\left(\frac{c_s^2 I_{4,0}-I_{4,1}}{I_{3,0}}\right)\right.\right.\nonumber\\&\left.\left.-\left(\frac{c_s^2 I_{r+2,0}-I_{r+2,1}}{I_{r,0}}\right)\right\}\right]\,,
    \label{Fr}
    \end{align}
and
\begin{align}
    \mathcal{H}_r&=\left[\frac{\left( \epsilon_0 + P_0 \right)}{\tau}\left[1+(b_0 + c_s^2)\frac{\tau_\mathrm{R}}{\tau}\right]\right.\nonumber\\&\left. -\frac{\tau_\mathrm{R} }{\tau^2}(b_0\epsilon_0+3\beta I_{32})\right]\,.
    \label{Hr}
\end{align} 
For a fixed set of $r$ values ($r=\{0,2,4\}$), the proper-time evolution of temperature  is obtained by numerically solving Eq.~\eqref{Teq_full} with a set of initial conditions corresponding to  relativistic heavy-ion collisions, namely, the initial temperature is considered to be $T_0= 0.5$~GeV at initial proper-time $\tau_0=0.5$~fm, the  relaxation time is taken as $\tau_R=0.5$~fm and the mass of the medium constituents is assumed to be temperature independent with a fixed value $m=0.3$~GeV. With these given initial conditions, the corresponding proper-time evolution of the bulk viscous pressure is then obtained from Eqs.~\eqref{H-Pi} and \eqref{RH_zeta} as shown in Fig.~\ref{Pi/P0}.

From the inset of Fig.~\ref{Pi/P0}, it can be observed that the temperature evolution is not sensitive to the choices of $r$. This can be understood by studying Eq.~\eqref{Teq_full} where we note that the evolution of temperature depends on the two coefficients $\mathcal{G}_r$ and $\mathcal{H}_r$. The expression of $\mathcal{G}_r$ in Eq.~\eqref{Gr} shows the $r$ dependent terms are suppressed by a factor of $\tau$ as compared to $I_{30}$. Similarly, we note from Eq.~\eqref{Hr} that the $r$ dependent terms are suppressed by a factor of $\tau$ here as well. Consequently, in the evolution of temperature, the variation due to $r$ is not very prominent. Rather the evolution is mostly dominated by the boost-invariant expansion of the system. On the other hand, the bulk viscous pressure, scaled by the equilibrium pressure, depends significantly on the choices of $r$. This is evident from the fact that in Eq.~\eqref{RH_zeta}, the $r$ dependent terms are on equal footing with the $r$-independent terms.

\section{Summary and outlook}
\label{sec:summ_conc}

In this work, we have provided the first formulation of relativistic dissipative hydrodynamics from BGK collision kernel, which represents a generalization of RTA collision kernel. We found that unlike RTA that does not allow the out of equilibrium corrections to energy and number density, in case of BGK, these corrections can be nonzero provided they satisfy Eq.~\eqref{BGK_MC_1}. While this is still not the most general scenario, it allows us to specify the coefficient of bulk viscosity, independently from shear viscosity, as opposed to RTA. We found that relativistic BGK hydrodynamics is controlled by a free parameter related to the freedom of a matching condition, which modifies the coefficient of bulk viscous pressure. On the other hand, the BGK kernel is ill defined for vanishing chemical potential as well as for a system with multiple conserved charges. We thus proposed a modified BGK collision kernel, which is free from such issues, and advocate it to be better suited for derivation of hydrodynamic equations. 

It is important to note that the BGK or MBGK collision kernels are affected by the matching conditions, which in turn affects the dissipative processes in the system. Moreover, at finite chemical potential, two descriptions become identical. We identified a class of matching conditions for which the homogeneous part of the solution to the relativistic Boltzmann equation vanishes, and RTA turns out to be a special case of that. We examined the effect of choice of matching condition on dissipative coefficients and also studied scaling properties of the ratio of coefficients of bulk viscosity to shear viscosity on the conformality measure. The importance of the bulk viscosity in the hydrodynamic evolution of quark gluon plasma has been emphasized in Refs.~\cite{Ryu:2015vwa, Ryu:2017qzn, Paquet:2015lta}. Our framework provides a direct control over this first-order transport coefficient through a choice of matching condition via $b_0$.

The present work with MBGK collision kernel demonstrates its advantages over RTA collision kernel in several ways. The formulation of hydrodynamics in the case of RTA collision kernel restricts the choice of frame and matching conditions to Landau choice. On the other hand, we have shown that MBGK represents a general class of collision kernels, with RTA being a specific case. This free parameter, associated with the choice of a matching condition, can be regulated to consider more realistic scenarios that may not be captured by the restrictive form of RTA. For instance, in the case of RTA, fixing the shear viscosity by specifying the relaxation time, completely fixes the bulk viscosity as well. Whereas in MBGK, the presence of the free parameter in the scalar dissipation  allows us to fix the bulk viscosity independently from shear viscosity. Moreover, by generalizing the collision kernel from RTA to MBGK, we have shown that we do not severely compromise on the simplicity of the form of the collision kernel necessary for hydrodynamic formulation.

The present formulation of hydrodynamics with a modified BGK collision kernel opens up several possibilities for future investigations. This MBGK collision kernel may also find potential applications in non-relativistic physics domain where BGK collision is widely used. The formulation of causal hydrodynamics with MBGK collision kernel is an immediate possible extension. Formulation of higher-order hydrodynamic theories may be affected more significantly as the evolution equations of scalar, vector, and tensor dissipative quantities contain cross-terms giving rise to the possibility of them being controlled by the matching conditions. Higher-order theories also exhibit interesting features like fixed points and attractors \cite{Heller:2015dha, Blaizot:2017ucy}, which could also be studied within the MBGK hydrodynamics framework. The present article forms the basis for all these studies which we leave for future explorations.

\section*{Acknowledgements}

The authors acknowledge Sunil Jaiswal for several useful discussions.  A.J. was supported in part by the DST-INSPIRE faculty award under Grant No. DST/INSPIRE/04/2017/000038. S.B. kindly acknowledges the support of the Faculty of Physics, Astronomy and Applied Computer Science, Jagiellonian University Grant No. LM/17/BS.
PS is supported by the European Union - NextGenerationEU through the grant No. 760079/23.05.2023, funded by the Romanian ministry of research, innovation and digitalization through Romania’s National Recovery and Resilience Plan, call no. PNRR-III-C9-2022-I8.

\bibliography{ref}

\end{document}